# The Lorenz number in the optimally doped and underdoped Eu-123 superconductor

M. Matusiak [1,*] and Th. Wolf [2]

[1] *Institute of Low Temperature and Structure Research, Polish Academy of Sciences,*

*P.O. Box 1410, 50-950 Wrocław, Poland*

[2] *Forschungszentrum Karlsruhe, Institut für Festkörperphysik, 76021 Karlsruhe, Germany*

The temperature dependences of the Hall-Lorenz number ($L_{xy}$) in a $EuBa_2Cu_3O_y$ (Eu-123) single crystal before and after oxygen reduction are reported. The study is based on data on the normal state longitudinal and transversal transport coefficients. Namely, the temperature dependences of the electrical resistivity, Hall coefficient, longitudinal thermal conductivity, and transverse thermal conductivity are presented. The set of measurements was performed for an optimally doped sample ($y \approx 7$), then the oxygen content in the same crystal was reduced to $y \approx 6.65$, and the measurements were repeated. For both cases $L_{xy}$'s are about two times larger than the Sommerfeld's value of the Lorenz number and depend weakly on temperature in a range 300 – 160 K. Below $T \approx 160$ K the Hall-Lorenz number for the optimally doped sample slowly drops, while the value of $L_{xy}$ for the oxygen reduced sample begins to rise. Such results suggest that the electronic system in the investigated compound may be considered as a metallic one with pseudo-gaps that open at the Fermi level.

---

[*] Corresponding author. Tel.: +48-71-3435021; Fax: +48-71-3441029; e-mail: M.Matusiak@int.pan.wroc.pl

A relation between thermal and electrical conductivities known as the Wiedemann-Franz (WF) law is a sensitive probe of fundamental properties of the electronic system. The value and temperature dependence of the Lorenz number (*L*), defined in the WF law as:

$$L \equiv \frac{\kappa_{el}}{\sigma T}\left(\frac{e}{k_B}\right)^2, \qquad (1)$$

where $\kappa_{el}$ is the electronic thermal conductivity, $\sigma$ is the electrical conductivity, $k_B$ is Boltzmann's constant, can bear testimony to the nature of the ground state of the electronic system. In the standard Fermi liquid theory *L* is equal to the Sommerfeld's value $L_0 = \pi^2/3$ as far as the mean free paths for transport of charge and entropy are identical. Deviations from the WF law may be regarded as a mark of an exotic nature of the electronic ground state. For example spin-charge separation[1,2], i.e. fractionalization of electrons for neutral spin-1/2 fermions called "spinons" and spinless charge-*e* bosons called "holons", is considered as a reason of violation of the WF law in a cooper oxide superconductor $(Pr,Ce)_2CuO_4$[3]. If charge carriers are bipolarons the Lorenz number will be strongly suppressed ($L \approx 0.15\ L_0$) and approximately proportional to the temperature[4,5]. On the other hand, results derived in the framework of the marginal Fermi-liquid approximation show *L* to be temperature independent, but still smaller than Sommerfeld's value ($L \approx 0.68\ L_0$)[6].

Unfortunately, the pure electronic Lorenz number can not be experimentally evaluated on the basis of measurements of the longitudinal transport coefficients, since heat in solids is usually carried by both electrons and phonons. Thus the measured total thermal conductivity is the sum of the electronic ($\kappa_{el}$) and phonon ($\kappa_{ph}$) components, $\kappa = \kappa_{el} + \kappa_{ph}$, where in high-$T_c$ cuprates $\kappa_{ph}$ provides a significant contribution to $\kappa$ and can not be neglected[7,8].

In this communication we use a unique method of a separation of the electronic component from the total thermal conductivity as proposed by Zhang et al.[9]. It is based on measurements of the transverse thermal and electrical conductivities ($\kappa_{xy}$ and $\sigma_{xy}$), since these coefficients appear to be purely electronic[9,10,11]. Thus we study the longitudinal and transverse transport effects in the optimally, and underdoped Eu-123 superconductor and present results on the determination of the Hall-Lorenz number ($L_{xy}$). Our studies have been performed in the normal-state of a $EuBa_2Cu_3O_7$ and $EuBa_2Cu_3O_{6.65}$ single crystal. The Hall-Lorenz numbers for both samples behave almost the same in a temperature range 300 – 160 K, but below $T \approx 160$ K the $L_{xy}$ coefficient for the optimally doped sample slowly drops continuously, while the value of $L_{xy}$ for the underdoped sample begins to rise. The values of both Hall-Lorenz numbers significantly exceed the Sommerfeld value. To explain the properties of the examined compound we suppose that a pseudo-gap opens at the Fermi level in the optimally doped sample, while two different pseudo-gaps appear in two temperature regions for the oxygen reduced crystal.

Single crystals of $EuBa_2Cu_3O_y$ were grown from CuO-BaO flux in $ZrO_2$ crucibles by the slow cooling method[12]. To avoid any substitution of Ba by Eu the growth was carried out in a reduced atmosphere of 100 mbar air. The crystal selected for measurements was about 2 mm long, 1.1 mm wide and 0.18 mm thick (along the crystallographic *c*-axis). It was twinned, therefore the coefficients measured along the *ab*-plane should represent values averaged between the *a* and *b* crystallographic directions. The magnetically measured superconducting transition of the sample had an onset of ≈ 94.5 K and a midpoint of 94.0 K. The oxygen content *y* was estimated to be close to 7 (on the basis of the $T_c$ value and the in-plane thermoelectric power value at room temperature $S_{ab}(300 K) \approx -2 \mu V/K$[13,14,15]).

As described below a second set of measurements was performed on the same specimen annealed in air for 12 days at $T = 500\,°C$ to reduce its oxygen content. After the reduction treatment the critical temperature ($T_c$) dropped to about 61 K (measured magnetically) and its thermoelectric power was equal to $S_{ab}(300\text{ K}) \approx 30\,\mu$V/K. Both values suggest[13,14,15] that the oxygen content has dropped to $\approx 6.65$.

The experimental techniques for measuring the electrical resistivity, Hall coefficient, thermal longitudinal and transverse conductivities were the same as described elsewhere[10,11].

Our studies are based on an analysis of the longitudinal and the transverse transport coefficients. The longitudinal thermal and electrical conductivities ($\kappa_{xx}$ and $\sigma_{xx}$) were measured in absence of a magnetic field, while for the measurement of the transverse thermal and electrical conductivities ($\kappa_{xy}$ and $\sigma_{xy}$) a magnetic field was applied. It is worth to underline that the object of investigations was one and the same $EuBa_2Cu_3O_y$ (Eu-123) single crystal, which was examined before and after oxygen reduction. In Figure 1 we show the temperature dependences of the electrical coefficients: $\sigma_{xx}(T)$ and $\sigma_{xy}(T)$. The longitudinal electrical conductivity is presented as the electrical resistivity ($\rho = 1/\sigma_{xx}$) in the upper panel (*a*). The transverse electrical conductivity is shown in the inset as a result of multiplication the Hall coefficient ($R_H$) by the square of $\sigma_{xx}$ ($\sigma_{xy}/B = R_H(\sigma_{xx})^2$). It is the valid equation, since the crystal is twinned, thus $\sigma_{xx} = \sigma_{yy}$. The temperature dependence of $R_H$ is presented in the bottom panel (*b*). As mentioned above, all curves relate to the same Eu-123 single crystal: at the beginning it was an optimally doped (solid lines), but after annealing in air it became underdoped (dashed lines). As expected, the electrical resistivity rises (about three times) at room temperature in the oxygen reduced sample. The critical temperature, defined as a point where curvature of $\rho(T)$ changes sign, drops from 94.5 K (optimally doped) to 62.5 K (underdoped). The Hall coefficient in the underdoped sample is about six times larger (at $T = 300$ K) than it was before oxygen reduction. However, the temperature dependences of $R_H$ and $\rho$ for the oxygen reduced sample still have the same characteristics as those observed for the optimally doped one.

There also no qualitative differences between thermal conductivities presented in Figure 2. For both samples the longitudinal thermal conductivity in the normal state slowly rises when the temperature decreases – see panel (*a*). $\kappa_{xx}$ for reduced sample is almost two times smaller (at room temperature) than it was for the optimally one. Also the transverse thermal conductivity $\kappa_{xy}$ is lower in reduced sample – by a factor of two at room temperature. The $\kappa_{xy}(T)$ dependencies for both samples can be fitted by a power function $aT^b$, where the parameter $b$ is equal to -1.14 for the optimally doped, and -1.21 for the underdoped sample. These values agree with results of measurements of the transverse thermal conductivity in an optimally doped $YBa_2Cu_3O_y$ single crystal performed by Zhang et al.[9] ($b = 1.21$), and independently by Matusiak et al.[11] ($b = -1.19$).

The above presented coefficients can be used to calculate the Hall-Lorenz number from the modified WF law[9]:

$$L_{xy} \equiv \frac{\kappa_{xy}}{\sigma_{xy}T}\left(\frac{e}{k_B}\right)^2. \qquad (2)$$

Results of evaluation of the Hall-Lorenz numbers for optimally doped (solid points) and oxygen reduced (open points) samples are presented in Figure 3. It is noticeable that both curves are practically the same above $T \approx 160$ K. Moreover, the data are very close to those observed in optimally doped $YBa_2Cu_3O_y$ single crystal[11]. The differences between our results and those reported for $YBa_2Cu_3O_y$ in Ref. 9 are discussed in Ref. 6 and 11. Below $T \approx 160$ K the Hall-Lorenz number for optimally doped sample decreases slowly, while $L_{xy}(T)$ in the oxygen reduced sample increases to significantly higher values.

Such a change of behavior has to mirror a change in the electronic structure, since the Hall-Lorenz number $L_{xy}$ is a pure electronic factor – there are no phonon contributions to the transverse thermal and charge conductivities. In addition, the Hall-Lorenz number is equal to the regular Lorenz number if charge carriers scatter elastically[16]. Otherwise, i.e. when inelastic scatterings appear, which usually disturb the heat current more effectively than the charge current[17], $L_{xy}$ may be compared to $L$ by the ratio:

$$a_L \equiv \frac{L^2}{L_{xy}}, \qquad (3)$$

The $a_L$ factor is expected to be nearly constant, since $L \sim \langle l_s \rangle / \langle l_e \rangle$ and $L_{xy} \sim \langle l_s^2 \rangle / \langle l_e^2 \rangle$[9,16], where ($l_s$) and ($l_e$) are different mean free paths defined for the transport of entropy and charge respectively. Because the values of the Hall-Lorenz numbers for both optimally doped and underdoped sample, tend to saturate at high temperatures we suppose that the room temperature is high enough to assume that charge carriers are scattered mostly elastically. Thus it is possible to write $L_{xy}(300) \approx L(300)$ and calculate the Lorenz numbers from Eq. 3. The temperature dependences of $L$ for both specimens are shown in Figure 4. Remarkable are the large values of $L$'s at room temperature, which are about two times larger than Sommerfeld's value of the Lorenz number expected for a regular Fermi-liquid system: $L_0 = \pi^2/3$. It would be consistent with experimental studies of the thermal transport in other HTSC at temperatures lower than $T_c$ and in high magnetic fields, that also suggest that the Lorenz number can exceed the Sommerfeld's value[13,18]. One of possible reasons of the enhancement of the Lorenz number in 1-2-3 high-$T_c$ superconductors is a pseudo-gap that opens at the Fermi level, as it was indicated in Ref. [19]. A very similar consideration was carried out by J.F. Goff[20,21] as an explanation of the anomaly in the Lorenz number[22,23,24] of Chromium. J.F. Goff has suggested that the unusual $L(T)$ dependence in Cr is a result of an anomalous distribution of the electronic states about the Fermi level ($\varepsilon_F$). For the calculation the Klemens' Moments Method[25] was used, which is a way of treating the Boltzmann equation so that the resulting transport coefficients become integral functions of energy only. It is assumed that the same relaxation times describe both electrical and thermal transport, the phonon system is in equlibrium and the Fermi level is chosen as the reference energy. The transport coefficients are expressed using moments of the function $\sigma(\in)$ called specific conductivity[25], where $\in$ is the reduced energy ($\in = E/k_B T$). The $n$'th moment is defined as:

$$M_n = -\int_{-\infty}^{+\infty} \sigma(\in) \in^n \frac{\partial f^0}{\partial \in} d\in, \qquad (4)$$

where $f^0$ is the equilibrium Fermi-Dirac distribution function: $f^0(\in) = (e^\in + 1)^{-1}$.

In such an approach the electrical conductivity become: $\sigma = M_0$,

the electronic thermal conductivity: $\kappa_{el} = (k_B/e)^2 T(M_2 - M_1^2/M_0)$,

the thermoelectric power: $S = (k_B/e)T(M_1/M_0)$,

and the Lorenz number is equal to: $L = (k_B/e)^2 (M_2/M_0) - S^2$.

Since the relaxation time of individual scattering mechanisms may be separated for an energy and temperature dependence: $\tau^{-1}(E,T) = f(E)g(T)$, a reduced specific conductivity: $\sigma_r(E) = \sigma(E)/\sigma(0)$ [20] can be defined, which tends to be temperature independent. Thus the transport coefficients may be rewritten in terms of new integrals $G_n(T)$, where: $M_n = \sigma(0)G_n(T)$, and $G_n(T) = -\int_{-\infty}^{+\infty} \sigma_r(\epsilon) \epsilon^n \frac{\partial f^0}{\partial \epsilon} d\epsilon$. Now the calculated quantities become the products of two factors: $\sigma(0)$, containing the temperature dependences of the scattering mechanism and the density of states at the Fermi level, and $G_n(T)$ which gives the temperature dependence from anomalies in the energy dependence of $\sigma(\epsilon)$. If the retardation effect of the internal electric field on the thermal conductivity is neglected, (it is, at most, an effect of the order of magnitude of a few percent in our samples), it can be write:

$$L = (k_B/e)^2 (G_2/G_0). \qquad (4)$$

Then, to calculate an $L(T)$ dependence there has to be proposed a type of irregularity occurring in the electronic structure. J.F. Goff considered two models of electronic structures containing a depletion of the reduced specific conductivity about the Fermi level. They were supposed to be a parabolic-shaped well in a single band model[20] or a parabolic and BCS-like gaps in model of two electronic subsystems[21]. Such depletions of $\sigma(\epsilon)$ may be related to a decrease of the density of states, therefore it is a structure very similar to the pseudo-gap observed in high-$T_c$ superconductors[26,27]. We suggest this phenomenological approach can be successfully used to illustrate our experimental data.

The $L(T)$ data of the optimally doped sample, presented in Figure 4, has been fitted with a numerical calculated function $G_2(T)/G_0(T)$ (Eq. 4), which has two free parameters describing a parabolic-like pseudo-gap. A model of the electronic structure used in the calculation is shown in the inset (*a*) in Figure 4. Despite the shape of the well was arbitrarily chosen and it is assumed to be temperature independent we have obtained a satisfying agreement between the fit and the experimental data. Moreover, the estimated width of supposed pseudo-gap ($2\Delta = 170$ meV) is of the same order of magnitude as reported for so called "large pseudo-gap" in YBCO[28]. While in the optimally doped crystal we observe an influence of the large pseudo-gap only, the growth of the $L(T)$ value in the underdoped sample below the temperature $T^* \approx 160$ K suggests an opening of the second, narrower, pseudo-gap. An appearing of two distinct pseudo-gaps seems to be characteristic for the high-$T_c$ superconductors phenomenon and it has been observed in many experiments[26,27,28,29,30]. Thus to describe this feature we have assumed in our calculations that the "large pseudo-gap" is simply replaced by the "small pseudo-gap" at $T^* = 160$ K. Both pseudo-gaps are presented in the inset (*b*) in Figure 4. The estimated width of the gap ($2\Delta = 65$ meV) and the temperature where it opens are similar to those observed previously for "small pseudo-gap" in underdoped YBCO single crystals with $T_c \sim 60$ K[31].

In summary, we have investigated transport properties in the normal state of a $EuBa_2Cu_3O_y$ single crystal. The sample was examined in two different oxidation states: first as an optimally doped one ($y \approx 7$), next, after oxygen reduction treatment, as an underdoped one ($y \approx 6.7$). In both cases the sample was superconducting, where the critical temperatures were equal to 94.5 K (optimally doped) and 62.5 K (underdoped). The measured temperature dependences of the longitudinal and transverse electrical, as well as thermal conductivities have been used to calculate the temperature dependences of the Hall-Lorenz numbers ($L_{xy}$). For both samples the values of $L_{xy}$ are significantly enhanced – at room temperature they are about two times larger than the Sommerfeld value of the Lorenz number. The found temperature dependences are practically identical from room temperature down to $\sim 160$ K. Below this temperature $L_{xy}$ for the optimally doped crystal still decreases slowly, while the value of $L_{xy}$ for the underdoped sample rises. The $L_{xy}(T)$ dependences were used to find the temperature dependences of the regular Lorenz numbers ($L$). To explain their behavior a phenomenological model of the electronic structure was proposed. The main feature of this model is a depletion of the density of states present at the Fermi level. In other words, we believe there is seen an opening of one pseudo-gap for the optimally doped crystal, and two different pseudo-gaps for the underdoped one.

The authors are grateful to Dr. T. Plackowski, Dr. C. Sułkowski and Dr. A.J. Zaleski for cooperation.

**Figure captions:**

1. Temperature dependences of the electrical transport coefficients for the EuBa$_2$Cu$_3$O$_7$ (optimal – solid lines) and EuBa$_2$Cu$_3$O$_{6.65}$ (reduced – dashed lines) single crystals. The longitudinal electrical resistivities are presented in the upper panel (*a*), and the Hall coefficients are placed in the bottom panel (*b*). The inset shows the transverse electrical conductivities calculated as: $\sigma_{xy}/B = R_H (\sigma_{xx})^2$.

2. Temperature dependences of the thermal conductivities for the EuBa$_2$Cu$_3$O$_7$ (optimal – solid lines, solid points) and EuBa$_2$Cu$_3$O$_{6.65}$ (reduced – dashed lines, open points) single crystals. The longitudinal coefficients are presented in the upper panel (*a*), and the transverse coefficients are placed in the bottom panel (*b*); the dash-dot lines, which are drawn along $\kappa_{xy}$ points show functions $aT^b$ being best power fits with exponents equal to $b_{opt}$ = -1.14 and $b_{red}$ = -1.21.

3. Temperature dependences of the Hall-Lorenz numbers for the EuBa$_2$Cu$_3$O$_7$ (optimal – solid points) and EuBa$_2$Cu$_3$O$_{6.65}$ (reduced – open points) single crystals; the dashed lines are guides for the eye; the dash-dot line shows the Sommerfeld's value of the Lorenz number ($L_0 = \pi^2/3$).

4. Temperature dependences of the Lorenz numbers for the EuBa$_2$Cu$_3$O$_7$ (optimal – solid points) and EuBa$_2$Cu$_3$O$_{6.65}$ (reduced – open points) single crystals; the solid lines present results of the *L*'s calculations, which are based on the assumption that a depletion of the reduced electrical resistivity appears at the Fermi level. In the inset (*a*) the shape of the depletion assumed for EuBa$_2$Cu$_3$O$_7$ is shown, while in the inset (*b*) the shapes of two wells assumed in two temperature regions for EuBa$_2$Cu$_3$O$_7$ are presented – for *T* > *T**  it is the solid line, and for *T* ≤ *T** it is the dashed line.

**Figure 1.**

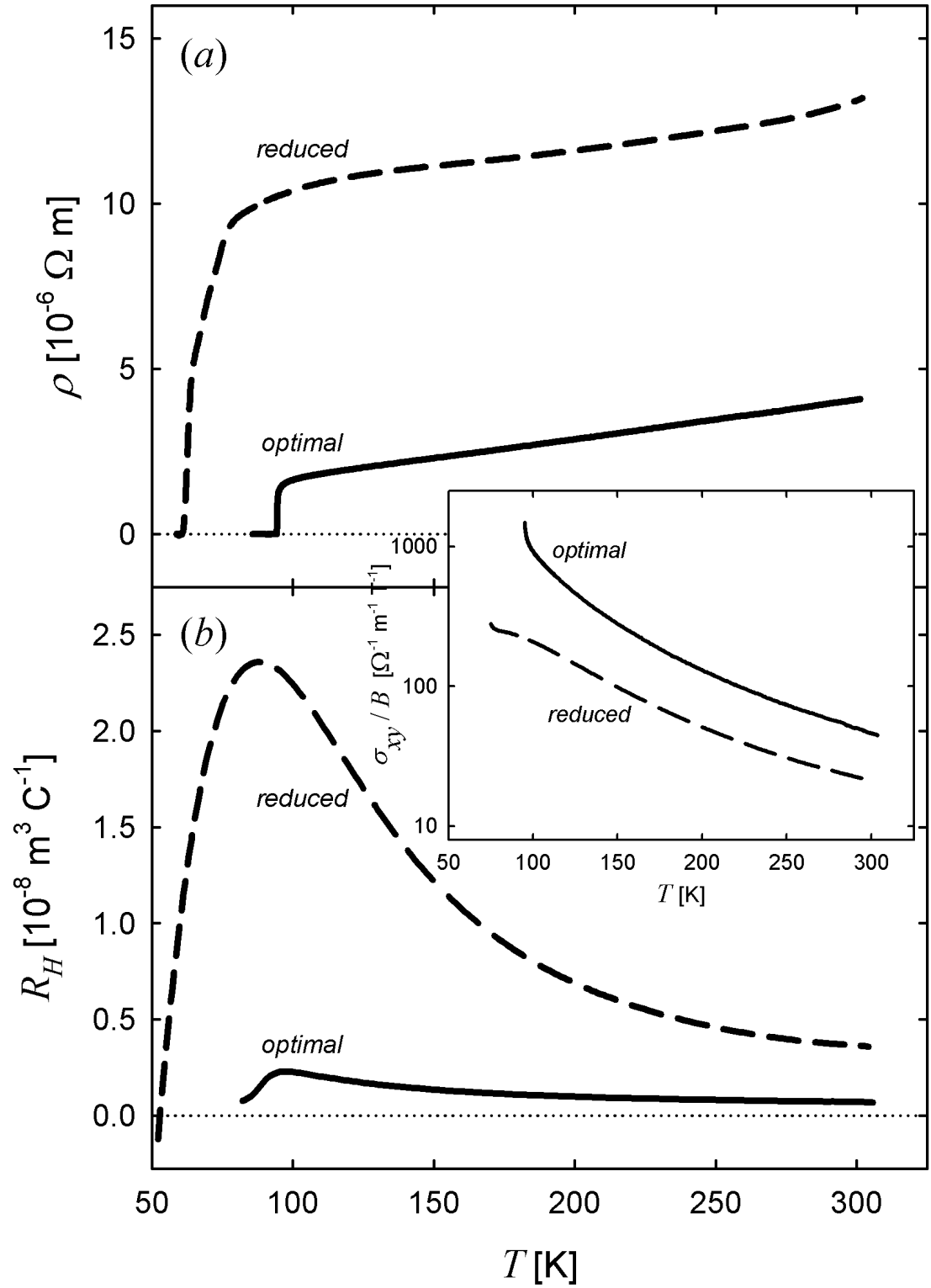

Figure 2.

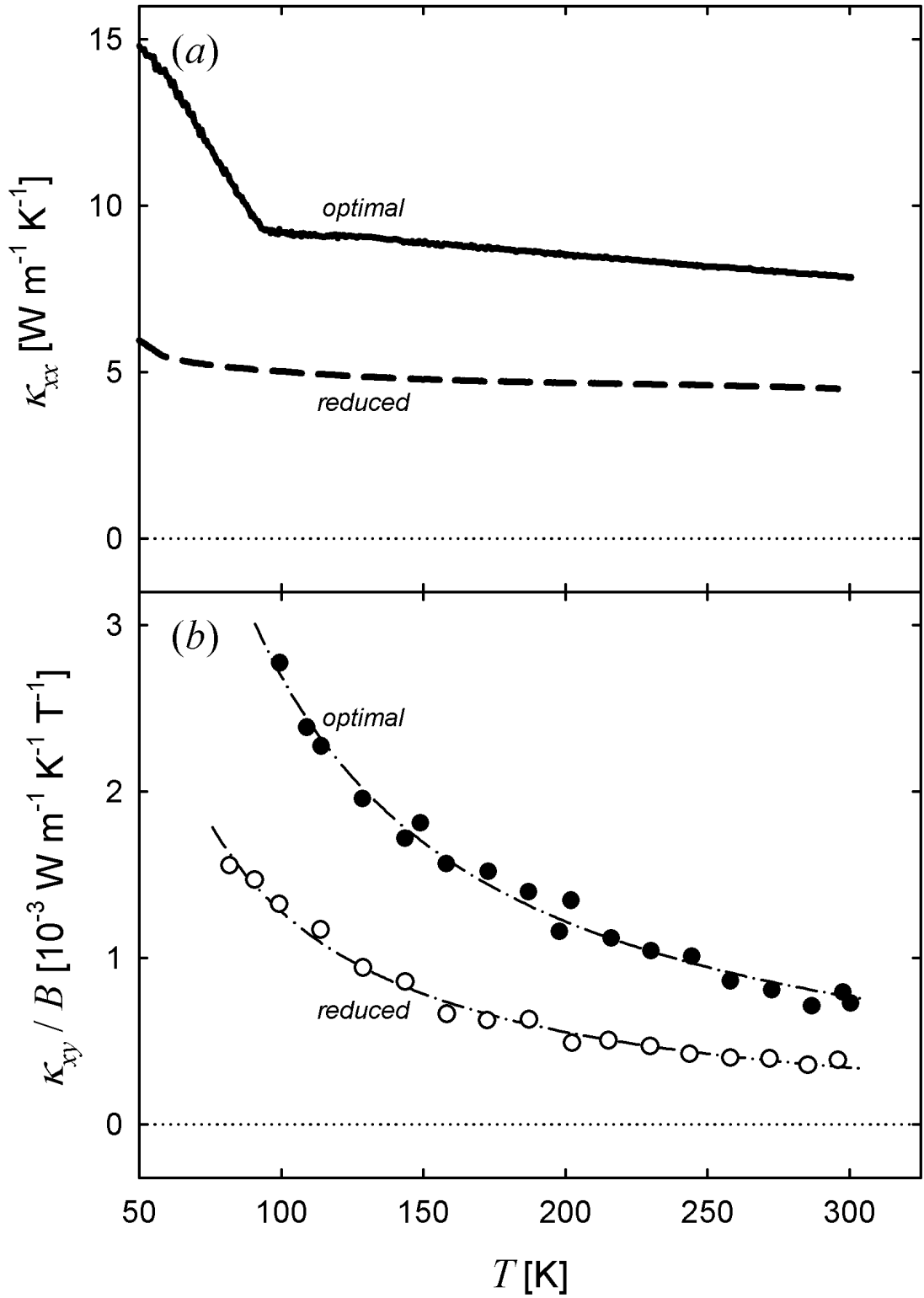

**Figure 3.**

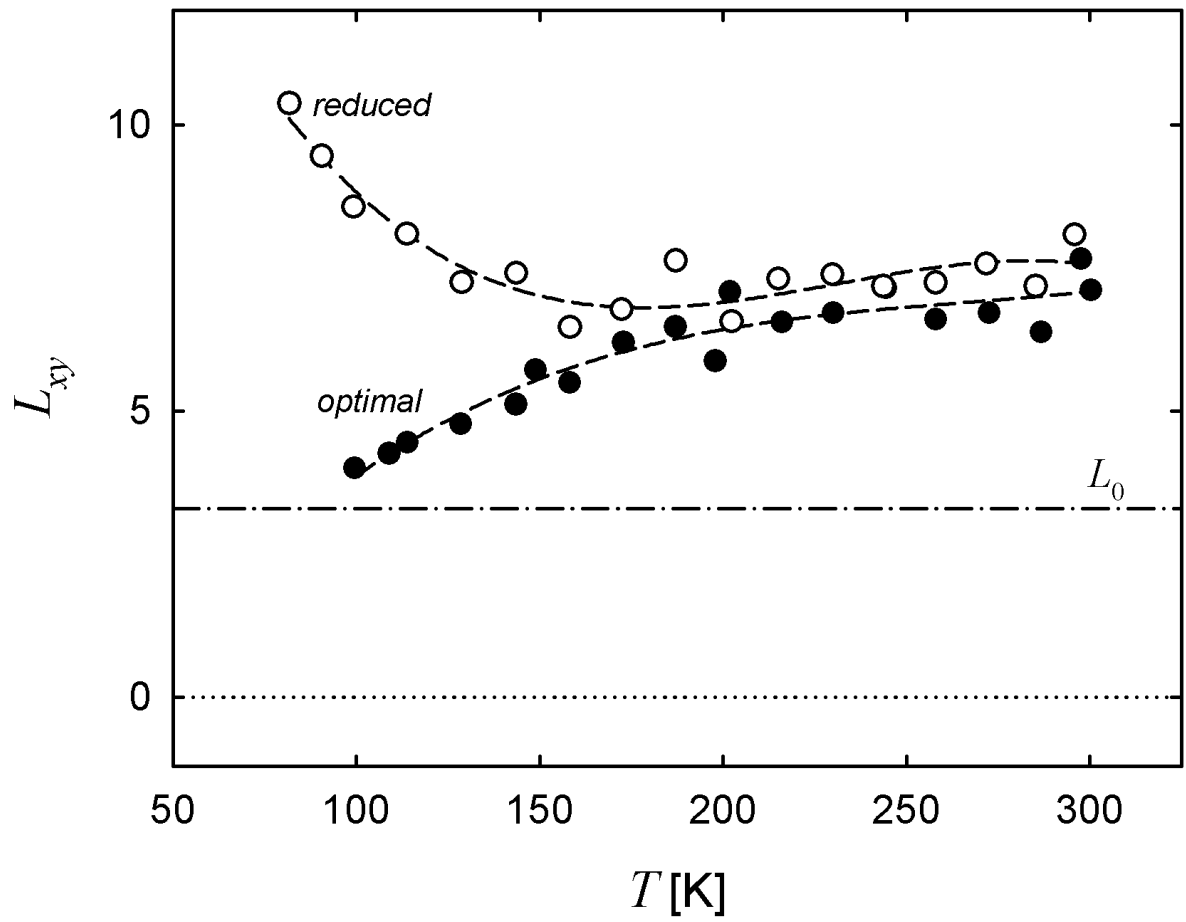

**Figure 4.**

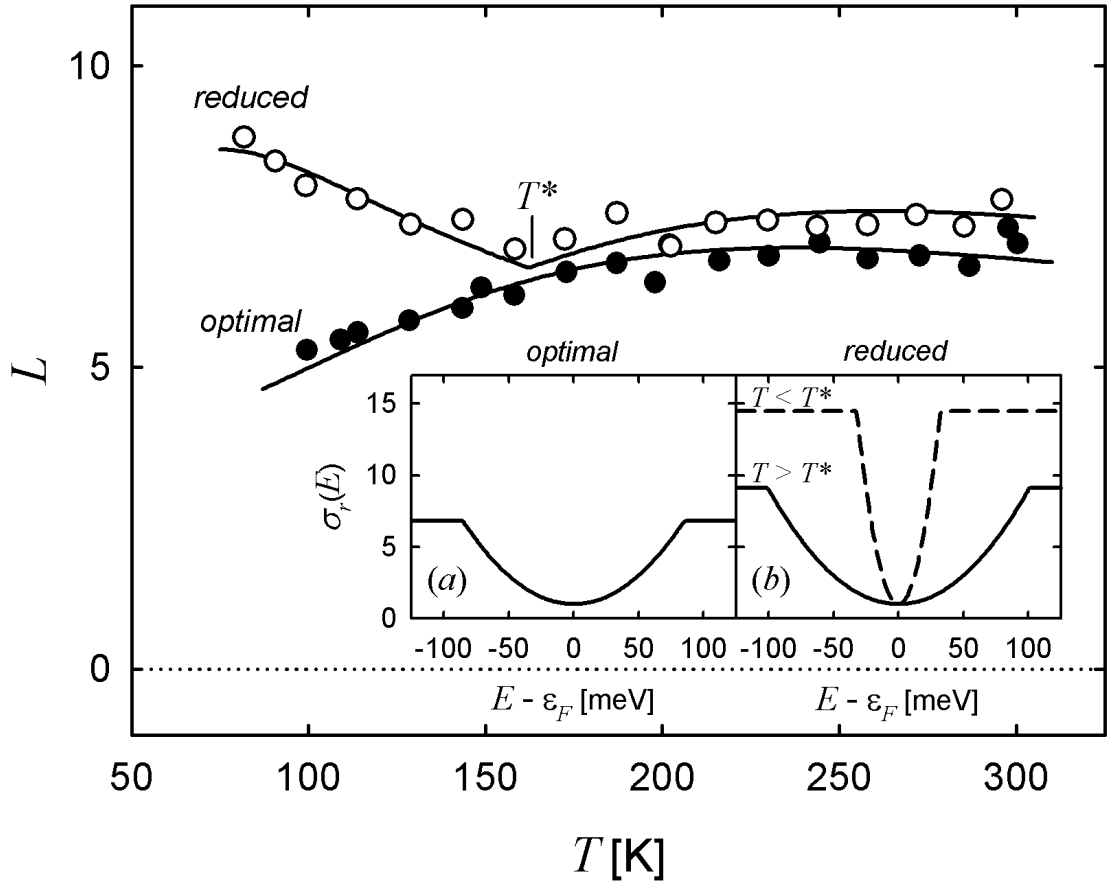